\documentclass[twocolumn]{aastex631}

\usepackage{here}
\usepackage[nooneline]{subfigure}
\usepackage{graphicx}

\shorttitle{The study of X-ray flux variability of M87}
\shortauthors{Imazawa et al.}

\begin{document}

\title{The Study of X-Ray Flux Variability of M87}

\author[0000-0002-0643-7946]{Ryo Imazawa}
\affiliation{Department of Physical Science, Hiroshima University \\
1-3-1  Kagamiyama, Higashi-Hiroshima, \\
Hiroshima 739-8526, Japan}

\author[0000-0002-0921-8837]{Yasushi Fukazawa}
\affiliation{Department of Physical Science, Hiroshima University \\
1-3-1  Kagamiyama, Higashi-Hiroshima, \\
Hiroshima 739-8526, Japan}

\author[0000-0001-6314-5897]{Hiromitsu Takahashi}
\affiliation{Department of Physical Science, Hiroshima University \\
1-3-1  Kagamiyama, Higashi-Hiroshima, \\
Hiroshima 739-8526, Japan}

\begin{abstract}
We searched for a short-term X-ray variability of the M87 core and jet from archival X-ray data with long exposure data taken by the Suzaku, Chandra, and NuSTAR telescopes.
We found the intraday variability for the Suzaku data obtained in 2006, and for the Chandra core obtained in 2017.
The intraday variability suggested a minute emission region of about the size of Schwartzshild radius of the M87 supermassive black hole.
Suzaku could not resolve a core and HST-1; however, in 2006, HST-1 was much brighter than the core, and thus, the variability is likely due to the HST-1.
Since the photon index in 2006 was 2.38, the emission was possibly synchrotron emission from the local shock region in the HST-1, indicating that the particle acceleration of TeV electrons occurred far away ($\sim$100 pc) from the core.
Assuming the fading time to be equal to the synchrotron cooling time, the magnetic field is constrained to be $B\sim$1.94 $\delta^{1/3}$ mG.
Moreover, the photon index of the core in 2017 was approximately 1.96; thus, the possible emission was from the radiative inefficiency accretion flow of the core or inverse Compton scattering in the jet.
Intraday time variability prefers the latter possibility.

\end{abstract}

\section{Introduction} \label{sec:intro}
\label{chap_intro}

A radio galaxy is an active galactic nucleus (AGN) with a strong jet. Some radio galaxies have been detected in the TeV gamma-ray band, indicating that particle  acceleration  up  to  TeV  energy  occurs.
The mechanism of particle acceleration up to TeV in the AGN is still under debate.
For example, \cite{axford1977} suggested a shock-in-jet model.
Intraday TeV gamma-ray variability found that IC 310 cannot be explained by the shock-in-jet model \citep{Aleksic1080}; thus, other models have been suggested, e.g., the minijet model \citep{Giannios09}, the cloud-jet interaction model \citep{Barkov10}, and the magnetospheric model \citep{BZ1977}.
Apart from the shock-in-jet model in the core jet region,
particle acceleration up to TeV energy in the outer jet region has been studied.
A recent discovery of extended TeV gamma-ray emission from the kpc-scale
jet of Cen A showed particle acceleration up to TeV in the kpc-scale jet region \citep{HESS2020}.

M87 is a nearby FR-I radio galaxy at $z$ $\sim$0.004 ($\sim$16 Mpc)\citep{Jordan_2005}, and jet components have been resolved through radio, optical, and X-ray observations.
Therefore, it is ideal to study the location of particle acceleration regions up to TeV, and many X-ray and gamma-ray observations have been performed.
In 2017, the Event Horizon Telescope team observed black hole shadow, and central black hole mass was estimated to be
$M_{\rm{BH}}$ = (6.5$\pm$0.7)$\times$ $10^{9}$ $M_\odot$ \citep{EHT2019}.
This mass is more massive than other known black holes \citep{Impey06}.
Gamma-ray emission has been detected in the GeV and TeV bands (\cite{Aharonian_2003}, \cite{Abdo09}).
Daily TeV gamma-ray variability was found by H.E.S.S \citep{aharonian} and MAGIC \citep{Albert_2008}, indicating that the emission region was as compact as the Schwarzschild radius of M87 supermassive black hole.
\cite{aharonian} suggested that daily TeV gamma-ray variability was due to particle acceleration at HST-1; $\sim$100 pc from the core.
Moreover, from later X-ray observations of the core and HST-1 flux variability, the core was suggested for the TeV emission region \citep{Acciari_2008}.
Therefore, whether HST-1 is a site of particle acceleration up to TeV energy or not is under debate.
In this article, to study the location of particle acceleration up to TeV, we searched for short-term X-ray variability from each jet component of M87 from archival X-ray data of Chandra, Suzaku, and NuSTAR.
If  we  find  a  rapid  variability  with  a  steep  X-ray  spectrum,  synchrotron emission by TeV electrons is suggested.

In this study, we assumed that the distance to M87 is 16 Mpc.
In Section 2, we describe observations and data analysis.
In Section 3, we describe the results of time variability and spectral analysis.
Finally, in Section 4, we discuss the origin of intraday X-ray time variability.\\

\section{Observation and Analysis}
\label{chap_obs}

To search for X-ray intraday variability,
we chose X-ray data that have a longer exposure time of more than 10 ksec.
For Chandra data, we chose data taken directly by Advanced CCD Imaging Spectrometer (ACIS) and High Resolution Camera (HRC); imaging observations without grating.
For ACIS data, we chose data taken with a short frame time of 0.4 sec (subarray mode) to minimize pile up effects.
As a result, 15 Chandra data (13 ACIS and 2 HRC), 6 NuSTAR data, and 1 Suzaku data were found.
There are 6 NuSTAR observations with more than 10 ksec, and Chandra observed simultaneously with each NuSTAR observation.
Since NuSTAR cannot resolve each jet component,
Chandra data was useful and thus these data are analyzed, even though their exposure time is short.
Tables \ref{tbl:obs}, \ref{tbl:obs2}, and \ref{tbl:obs3} summarize the list of observation data analyzed in this article.

We analyzed Chandra data with \texttt{CIAO}(version 4.12), and
Suzaku and NuSTAR data with \texttt{HEASoft v6.28}.
\texttt{CALDB} (version 20180925) were used as the calibration data for Suzaku and NuSTAR data analysis.
For Chandra data,
we checked pile-up effects in the 3$\times$3 pixels around the core in each observation, following the method of \cite{David2002}, and we confirmed that a pile-up fraction was $\leq 5\%$ in the core region for all observations analyzed here, except the observation (ID 21076) in the bright state in 2018, therefore the effects are negligible.
For this observation (ID 21076), the pile-up fraction was around 7.5\%.
Therefore this is not so significant for light curve analysis, but could affect the spectral fitting. and thus we note this issue in \S3.

\begin{table*}
    \centering
    \caption{List of Chandra data analyzed in this paper.
ID:13515 and ID:18612 were taken by HRC, marked with asterisk to the right of Obs ID.
Others data were taken by ACIS. 
Results of the short-term variability study are also presented as root mean square (RMS) (see Section 3). "-" means that variability was insignificant.}
        \smallskip
      \begin{tabular}{c c c c c}
        \hline
     Start Date (YYYY-MM-DD) & Obs ID & Exposure Time (ksec) & RMS & RMS$_{bgd}$\\\hline
2012-04-14 & 13515* & 74.3& - & -\\
2016-02-23 & 18233 & 37.2& 0.04$\pm$0.02 (core) & -\\
2016-02-24 & 18781 & 39.5& - & -\\
2016-02-26 & 18782 & 34.1& - & -\\
2016-04-20 & 18783 & 36.1& 0.05$\pm$0.03 (core) & 0.22$\pm$0.09\\
2016-04-28 & 18836 & 38.9& - & -\\
2016-05-28 & 18838 & 56.3& 0.06$\pm$0.02 (knot A) & -\\
2016-06-12 & 18856 & 25.5& - & -\\
2017-02-15 & 19457 & 4.6& - & -\\
2017-03-02 & 18612* & 72.5 & 0.07$\pm$0.01 (core)& -\\
&   &  & 0.09$\pm$0.02 (HST-1) & -\\
&       &      &0.04$\pm$0.02 (knot A)& -\\
2017-04-11 &  20034 & 13.1& - & -\\
2017-04-14 &  20035 & 13.1& 0.08$\pm$0.06 (knot A) & -\\
2018-04-24 &  21076 & 9.0& - & -\\
2019-03-27 &  21457 & 14.1& - & -\\
2019-03-28 &  21458 & 12.8& - & -\\
    \hline
  \end{tabular}\\
  \label{tbl:obs}
\end{table*}

\begin{table*}
    \centering
        \caption{List of NuSTAR data analyzed in this study.
For each ID, an upper ID is taken by Focal Plane Module-A (FPMA) and a lower is taken by FPMB.}
        \smallskip
      \begin{tabular}{c c c c c}
        \hline
     Start date (YYYY-MM-DD) & Obs ID & exposure time (ksec) & RMS & RMS$_{bgd}$\\\hline
     2017-02-15 & 60201016002 & 50.3 & 0.09$\pm$0.08  & -\\
& & & 0.04$\pm$0.04  & 0.04$\pm$0.02\\
              2017-04-11 & 90202052002 & 24.4 & - & -\\
&              &       & 0.04$\pm$0.04  & -\\
     2017-04-14 & 90202052004 & 22.5 & - & -\\
&              &       & - & - \\
              2018-04-24 & 60466002002 & 21.1 & - & -\\
&              &       & -  & -\\
     2019-03-26 & 60502010002 & 23.6 & - & -\\
     &              &       & - & -\\
              2019-03-28 & 60502010004 & 20.9 & - & -\\
&              &       & - & -\\  
    \hline
  \end{tabular}
  \label{tbl:obs2}
\end{table*}

\begin{table*}
    \centering
\caption{List of Suzaku data analyzed in this paper.}
        \smallskip
      \begin{tabular}{c c c c c}
        \hline
              Start date (YYYY-MM-DD) & Obs ID & exposure time (ksec) & RMS & RMS$_{bgd}$\\\hline
     2006-11-29 & 801038010 & 50.3 & 0.071$\pm$0.007 (XIS0) & -\\
     &    &      & 0.078$\pm$0.007 (XIS1) & -\\
     &    &      & 0.074$\pm$0.007 (XIS3) & 0.02$\pm$0.02\\

     \hline
  \end{tabular}
  \label{tbl:obs3}
\end{table*}
\subsection{Light curve}

\begin{figure}
 \centering
        \begin{tabular}{c}
                \subfigure[]{
                        \includegraphics[width=7cm,clip]{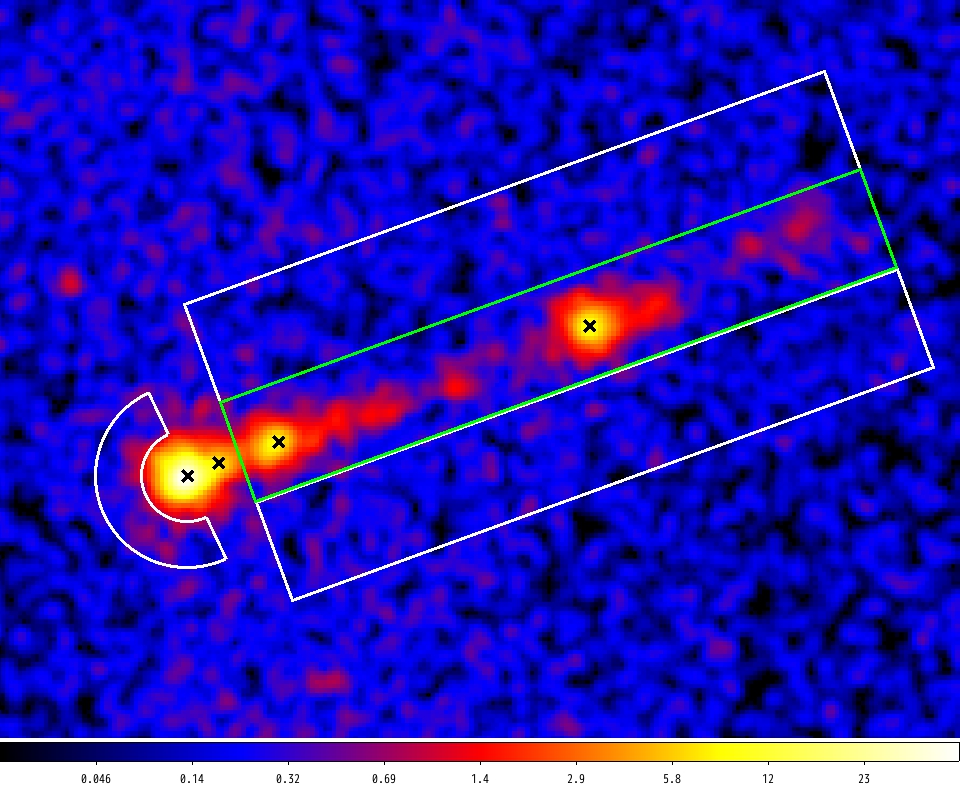}
                \label{fig:c2_picture}
                }\\
                \subfigure[]{
                        \includegraphics[width=7cm,clip]{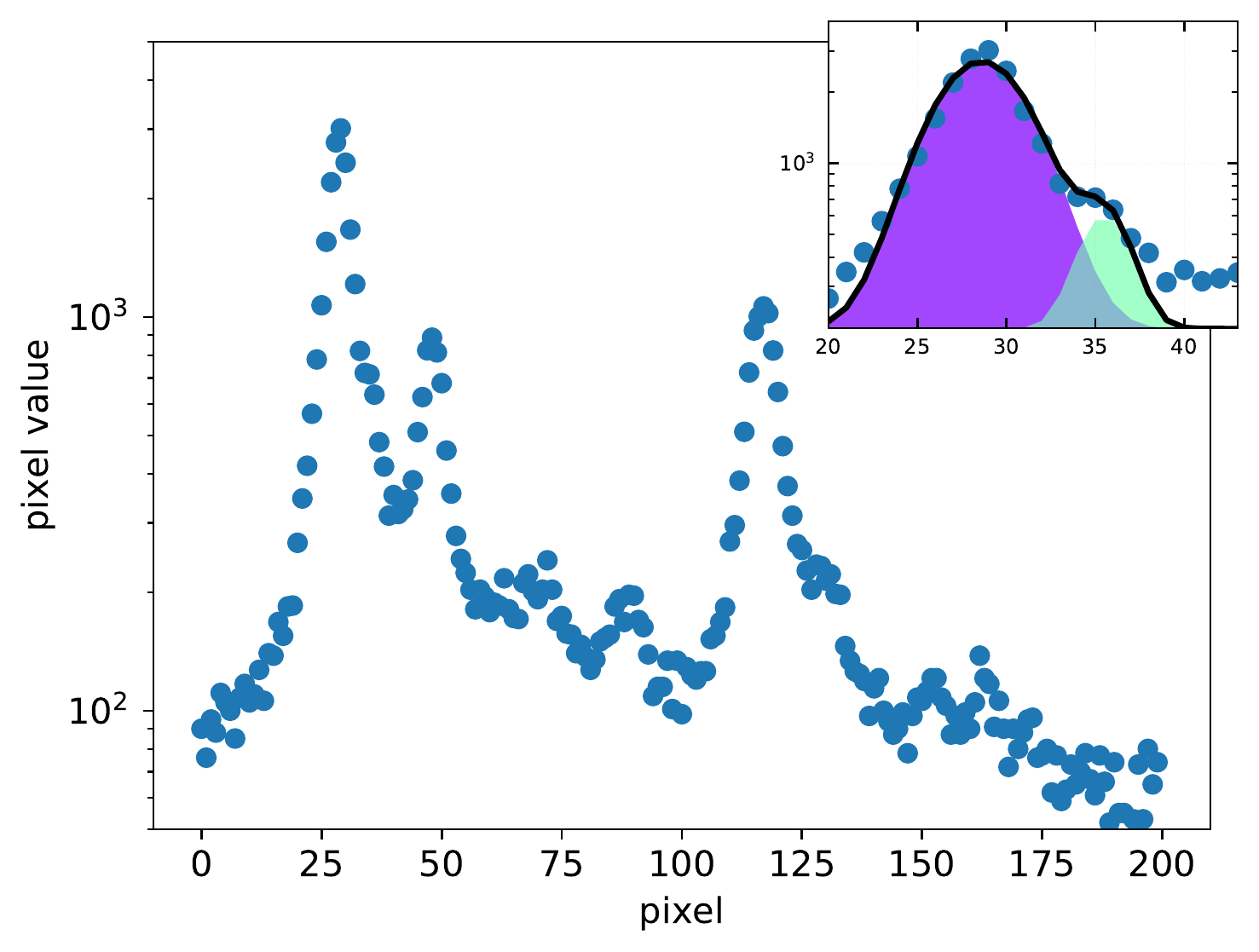}
                \label{fig:projection}
                }
                \\
\end{tabular}
        \caption{(a) M87 X-ray image obtained via Chandra/ACIS (2019-3-27; ID21457). Crosses in the figure show positions of the core, HST-1, knot D, and knot A from the left. The green box represents the jet region.
The white half-circular band shows the background region of the core and HST-1 and the white boxes show the background region of the jet (including knots D and A) for spectral analysis.
(b) Projected count rate profile for the R.A. direction obtained via Chandra HRC (2017-3-02; ID18612).
The top right portion figure is an enlarged view, which shows a two-Gaussian fit around the core (purple)  and  HST-1  (green).}
\label{fig:lc_fast}
\end{figure}

\begin{figure}
 \centering
 \includegraphics[width=9cm,clip]{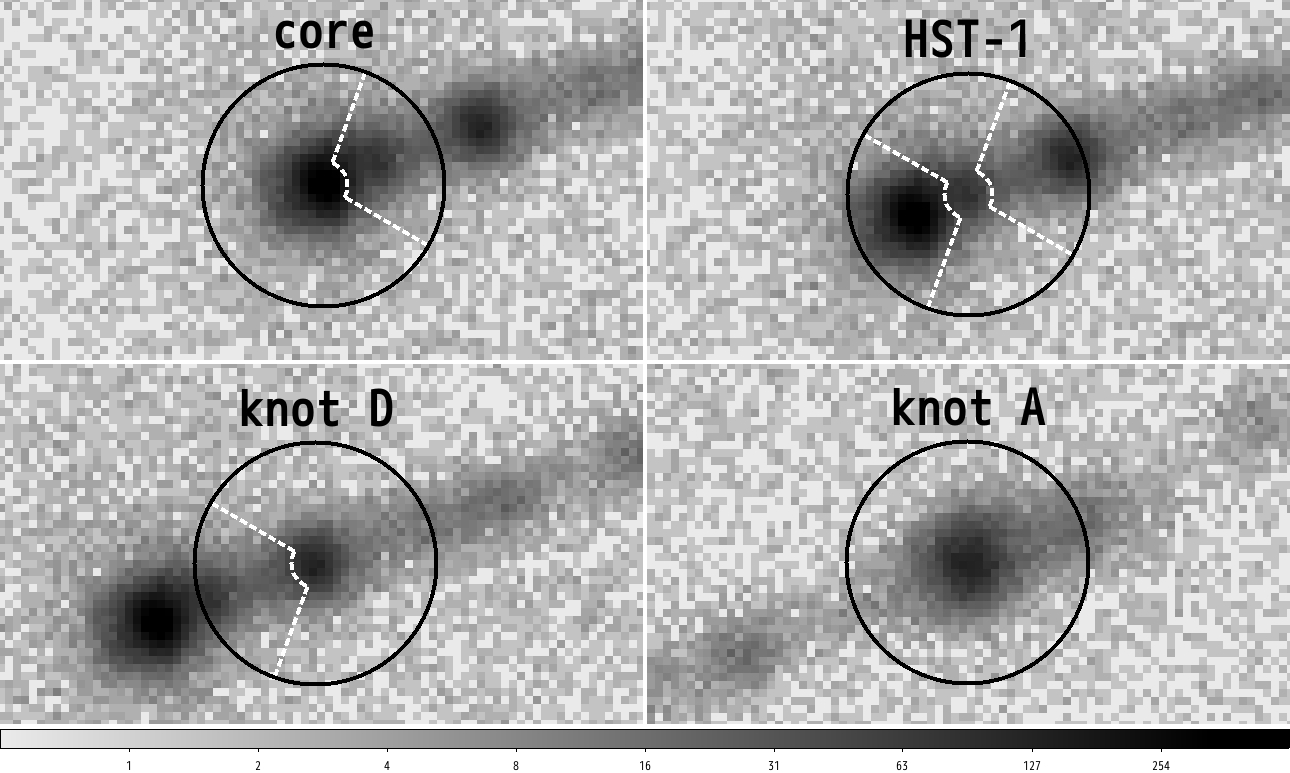}
        \caption{Extraction  regions  of  light  curves  of  each  component  in Chandra data.
Light curves are extracted in the black circular region, but the white dotted fan-shaped regions are  excluded.}
 \label{img:chan}
\end{figure}

To study a short-term variability, we derived an X-ray light curve in each observation.
For Chandra, we derived light curves of each component in the jet: core, HST-1, knot D, and knot A (see Figure \ref{fig:lc_fast}).
The extraction region of each component is shown in Figure \ref{img:chan}.
The position at 0.8 arcsec from the core is defined as HST-1. In addition,
knots D and A are defined in order from the core, and their positions are set at 2.9 and 12.4 arcsec from the core, respectively.
In each observation, the distance from the core for each  component  was  fixed,  and  the  coordinate  of  the  core was adjusted.
Each region is defined as a circle with a radius of 2.0 arcsec, but a fan-shaped region is excluded when the neighboring component exists in the circular region.
Most Chandra data were taken after 2016, when the core was brighter than HST-1.
Therefore, HST-1 signals were contaminated by the core signals.
To evaluate this effect, Figure \ref{fig:projection} shows a projected count rate profile along the right ascension (RA) direction.
The core and HST-1 positions are at 27 and 36 pixels, respectively.
The core emission is dominant, and a significant fraction of signals at HST-1 comes from the core.
Moreover, the effect of HST-1 on the core is insignificant and can be ignored.
Therefore, it is difficult to study the HST-1 variability but we will be able to mention the HST-1 variability if there is an independent variation with respect to the core.
Chandra light curves were created by the \texttt{dmextract} command in \texttt{CIAO} in the energy range of 2.0--10.0 keV (HRC data was not filtered) with a time bin of 3600 ksec.
To evaluate the variability of the background signal, light curves of the background region are created in the same way to source light curves, but the position is at 5.0 arcsec from the core in the direction opposite to the jet.

For NuSTAR, we derived light curves of FPMA and FPMB separately in each observation.
The source area is defined as a circle with a radius of 30 arcsec.
A light curve was created with the \texttt{lcurve} command in an energy range of 8.0--30.0 keV with a time bin of 3600 s.
As shown in $\S$2.2.1, the AGN emission is dominant above 8.0 keV.
The light curves of the background region were made with the same radius as the source region at 3.2 arcmin from M87.

The Suzaku satellite observed M87 for two days from  November 29 to December 1, 2006 (MJD: 54068-54070).
In this study, we used data obtained via X-ray Imaging Spectrometer (XIS).
XIS consists of three available CCDs in this observation, XIS0, XIS1, and XIS3.
XIS0 and XIS3 are front-illuminated CCDs, whereas XIS1 is a back-illuminated CCD.
Therefore, the sensitivity of the low energy range of XIS1 is superior to XIS0 and XIS3 \citep{XIS}.
In each instrument, a circle with a radius of 3.0 arcmin was defined as the source region.
The light curves of each CCD are created with the \texttt{lcurve} command in an energy range of 3.0--10.0 keV with a time bin of 8400 s.
Circles with a radius of 3.0 arcmin at 8.0 arcmin from the core were defined as background regions.

\subsection{Spectral analysis}
\subsubsection{NuSTAR and Chandra}
\label{sec:anachan}

\label{sec:nu}
There are six observations taken by both NuSTAR and Chandra simultaneously.
Following the method of \cite{Wong_2017}, we performed simultaneous fitting of NuSTAR and Chandra spectra to obtain the photon index and flux of each component.
In the simultaneous fitting, one  NuSTAR  spectrum  and  three  Chandra spectra of the core, HST-1, and jet were fitted with the model of hot gas emission were considered for the NuSTAR spectrum, and each power-law model was considered for the NuSTAR spectrum and one Chandra spectrum of each component.
Power-law parameters were common between NuSTAR and Chandra.

NuSTAR spectra were extracted in a circular region with a radius of 30 arcsec, and background regions were taken at 80--130 arcsec from M87. Spectra were binned so that 1 bin contained more than 20 photons.
For Chandra spectra of each component, the source region of core and HST-1 were defined as a circle with a position specified in Figure \ref{fig:c2_picture} and a radius of 0.4 arcsec.
For the background of core and HST-1 spectra, a semicircular band region at 2--4 arcsec was specified in the direction opposite to the jet (Figure \ref{fig:c2_picture}).
The jet area, including knots D and A, is defined as shown in Figure \ref{fig:c2_picture}, and the background region was specified as 19.5 arcsec $\times$ 3 arcsec boxes on both sides of the jet.
We created Chandra spectra with the \texttt{specextract} command and binned the spectra so that 1 bin contained more than 20 photons.

\begin{figure}[H]
 \centering
 \includegraphics[width=9cm,clip]{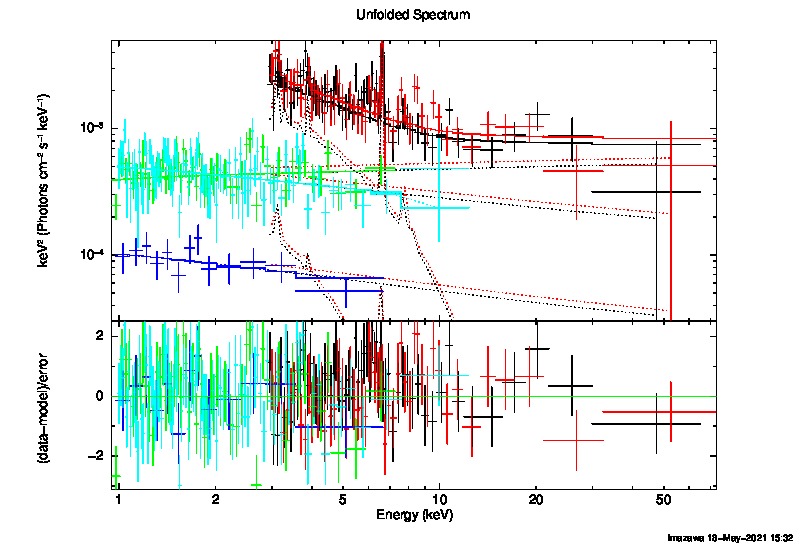}
        \caption{
An example of spectral fitting of one NuSTAR and three Chandra spectra obtained on 2019 March 28. The black and red data points represent the FPMA and FPMB of NuSTAR, respectively. The green, blue, and cyan points represent Chandra data points of the core, HST-1, and jet, respectively. The solid line is a total model spectrum, and dashed lines represent each spectral component. The bottom panel shows a residual.
}
\label{fig:nu_chan_spec}
\end{figure}

The spectral model for simultaneous fit consists of two temperature thermal plasma and three power-law models with Galactic absorption: \texttt{wabs*(vapec+vapec+pegpwrlw+pegpwrlw+pegpwrlw)} in \texttt{XSPEC}.
The normalization of \texttt{vapec} was set to 0 for Chandra spectra, and \texttt{vapec} parameters for NuSTAR were fixed as described later.
A column density of Galactic absorption was fixed to 1.94$\times10^{20} \rm{cm}^{-2}$ \citep{Kalberla_2005}.
Three  \texttt{pegpwrlw}  models  represent  emission  from  the  core,  HST-1, and  jet. 
Thus, for each Chandra spectrum of the core, HST-1, and jet, only one \texttt{pegpwrlw} normalization was left free and other normalizations were set to 0.
For these analyses, photon index and normalization were unified between NuSTAR and Chandra.
Considering cross-calibration is uncertain between FPMA and FPMB of NuSTAR, we added a constant model.
The constant was fixed to 1 for FPMA and Chandra spectra, whereas the constant was left to be free for FRMB.
The \texttt{vapec} parameters for NuSTAR spectra were determined as follows.
Since the extraction regions of the NuSTAR spectra are as wide as 30 arcsec, we must consider X-rays from the thermal plasma of clusters of galaxies.
However, since thermal emission is dominant in the soft X-ray band,
it is difficult to estimate them by only by the NuSTAR spectrum.
Therefore, first, we estimate the thermal emission using Chandra data.
Among observations of M87 by Chandra/ACIS, a full-array observation (2000-07-29; ID352) was selected.
We extracted a spectrum within 30 arcsec of M87, excluding the core and jet region.
For the background region, we specified a ring shape with an inner and outer radius of 80 and an 130 arcsec, respectively.
We fitted this spectrum with two \texttt{apec} models, following \cite{Russell15}, but instead of \texttt{apec} we used \texttt{vapec}.
A redshift was fixed to 0.0043 \citep{M87ned}.
The metal abundances of elements other than He, C, and N were free, but treated as common parameters between two \texttt{vapec} models.
He, C, and N abundances were fixed to 1 solar.
This model fits the spectrum well with a reduced chi-square of 1.28.
As a result, temperatures were determined as $kT$ = 1.66, 0.98 keV, and normalizations were determined as 5.10$\times10^{-3}$, 8.68$\times10^{-4}$.
Thus, the obtained parameters were used as fixed parameters in the simultaneous fitting of NuSTAR and Chandra spectra.
An example of simultaneous fittings is shown in Figure \ref{fig:nu_chan_spec}.
The spectra were fitted well.

For reference, we performed spectral fitting for only the NuSTAR spectrum with \texttt{wabs*(vapec+vapec+pegpwrlw)} and obtained a photon index and flux in 10.0--40.0 keV.

\subsection{Suzaku}
\begin{figure}[H]
 \centering
        \begin{tabular}{c}
          \subfigure[]{
\includegraphics[width=7cm,clip]{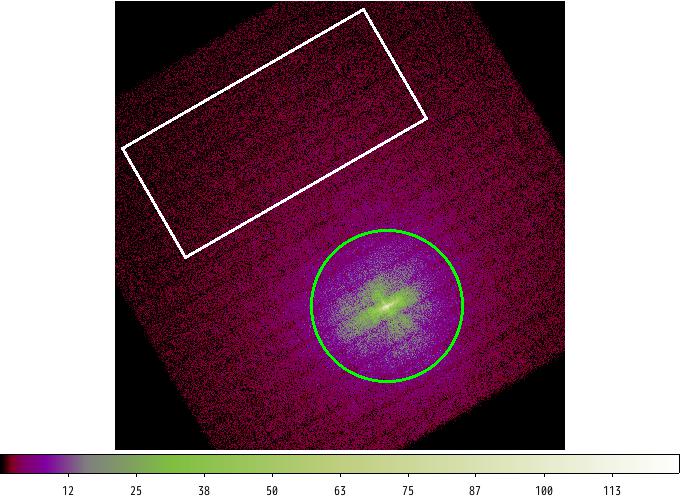}
\label{fig:suzakureg}
}\\
    \subfigure[]{
\includegraphics[width=7cm,clip]{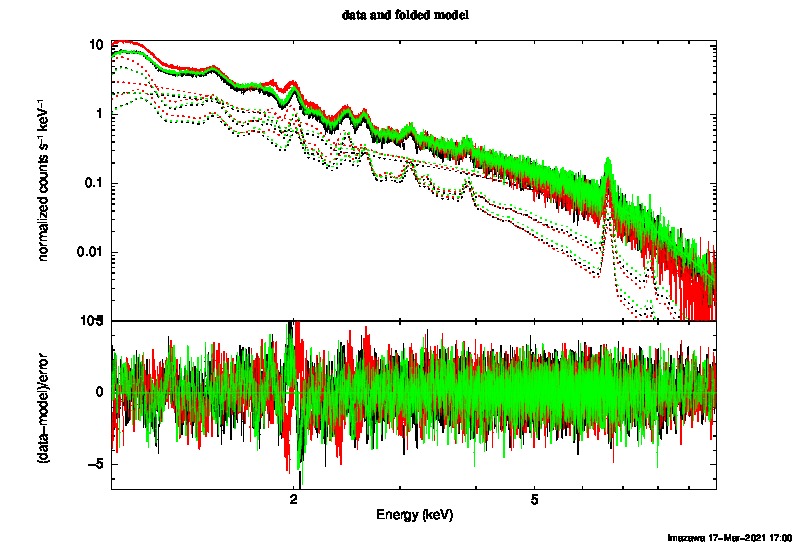}
        \label{fig:suzakuspec}
}\\
\end{tabular}
\caption{(a) Extraction region of Suzaku spectra overlaid on the XIS1 image.
The green circle with 3 arcmin is the source region, and the white  box  (5  arcmin$\times$11 arcmin) is the background region.
(b) Spectral fitting of Suzaku/XIS spectra. The black, red, and green data points represent XIS0, XIS1, XIS3, respectively.
 A solid line represents the total model spectrum, and the dashed lines represent each of spectral component. The bottom panel shows a residual.
}
\end{figure}

Spectra of Suzaku/XIS was extracted for the source region with a radius of 3 arcmin (Figure \ref{fig:suzakureg}).
Response files were made by the \texttt{xisarfgen} and\texttt{xisrmfgen} commands.\\

We  fitted  the  spectra  in  1.0--10.0  keV  with \texttt{wabs*(vapec+vapec+pegpwrlw)}.
The metal abundances of H, C, and N were fixed to 1, and the abundance parameters of other metals are unified between the two \texttt{vapec} models.
Figure \ref{fig:suzakuspec} shows a result of spectral fitting.
As a result, temperatures became 2.45 and 1.53 keV.
The spectra were well fitted with a reduced $\chi$ square of  $\chi^2 / \rm{d.o.f} = 1.25$.
The photon index became $\gamma = 2.38^{+0.07}_{-0.04}$ (2.0--10.0 keV), and the power-law flux became $2.71 ^{+0.18}_{-0.18}\times$ $10^{-11}$ erg cm$^{-2}$ s$^{-1}$.\\

\section{Result}
\label{chap_res}

\begin{figure*}
 \centering
        \begin{tabular}{c c}
               \subfigure[Chandra knot A (2016-05-28)]{
\includegraphics[width=7cm,clip]{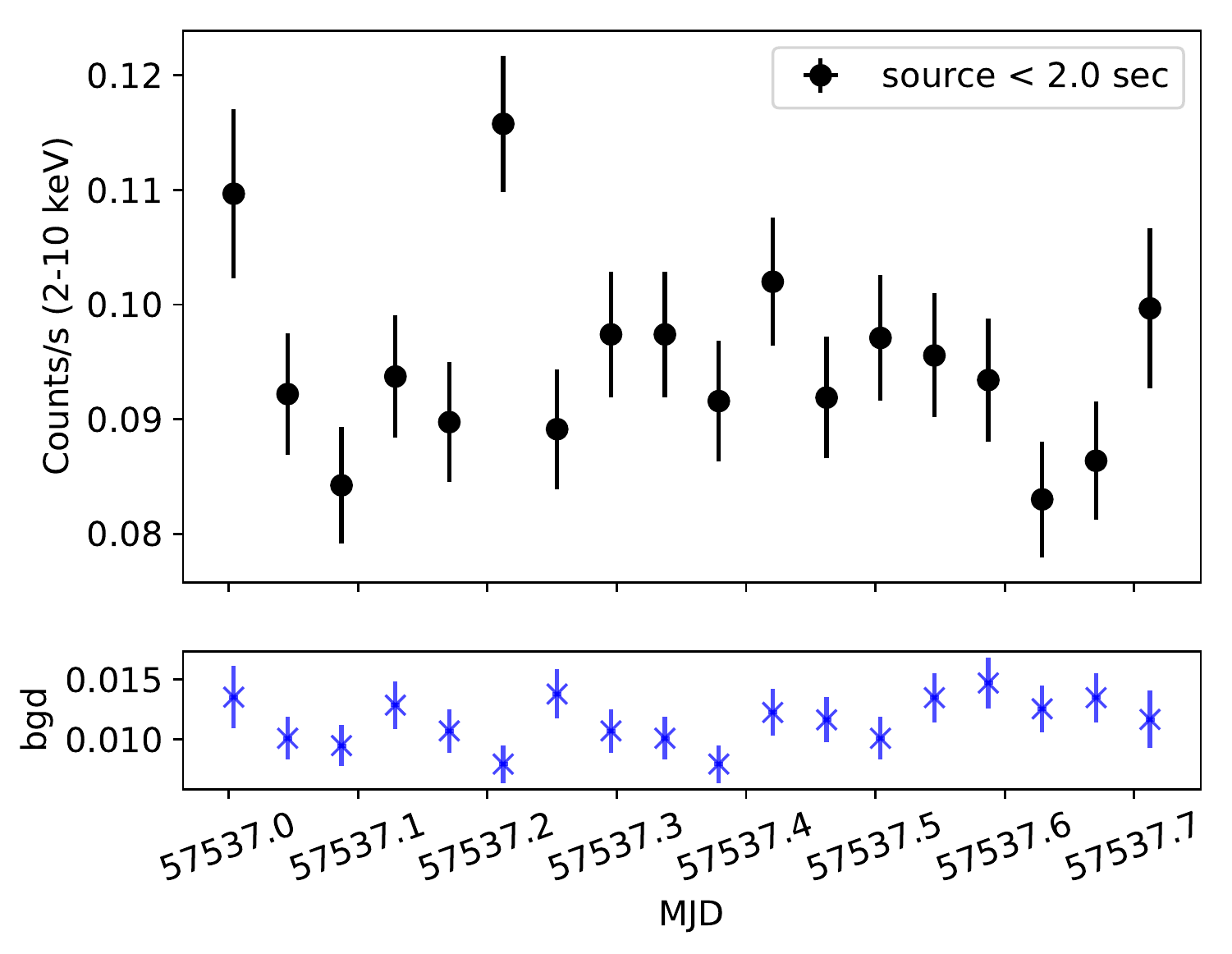}
        \label{fig:chan_2017_k1}}
               \subfigure[Chandra core (2017-03-02)]{
\includegraphics[width=7cm,clip]{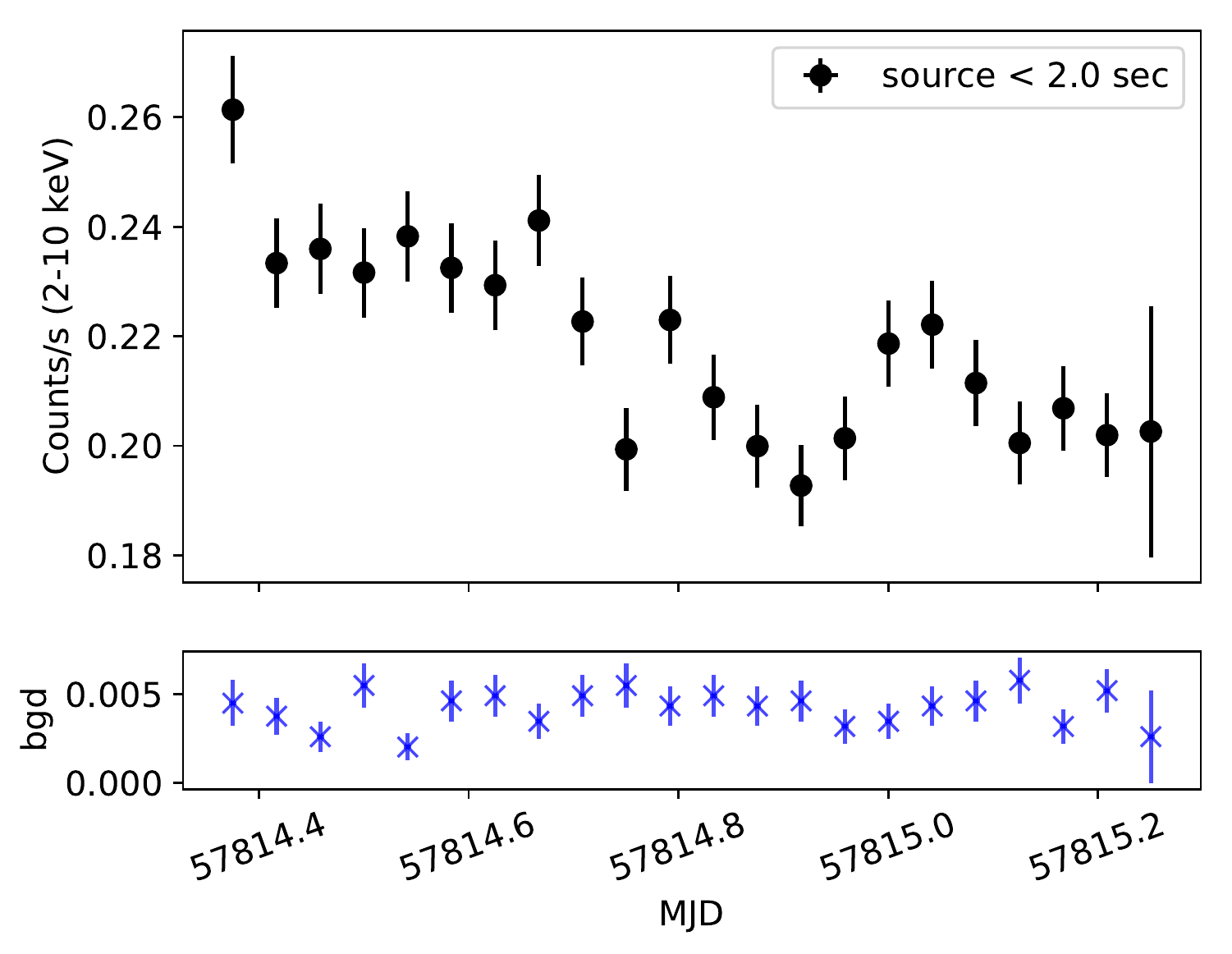}
\label{fig:chan_2017}}\\
                \subfigure[Chandra HST-1 (2017-03-02)]{
\includegraphics[width=7cm,clip]{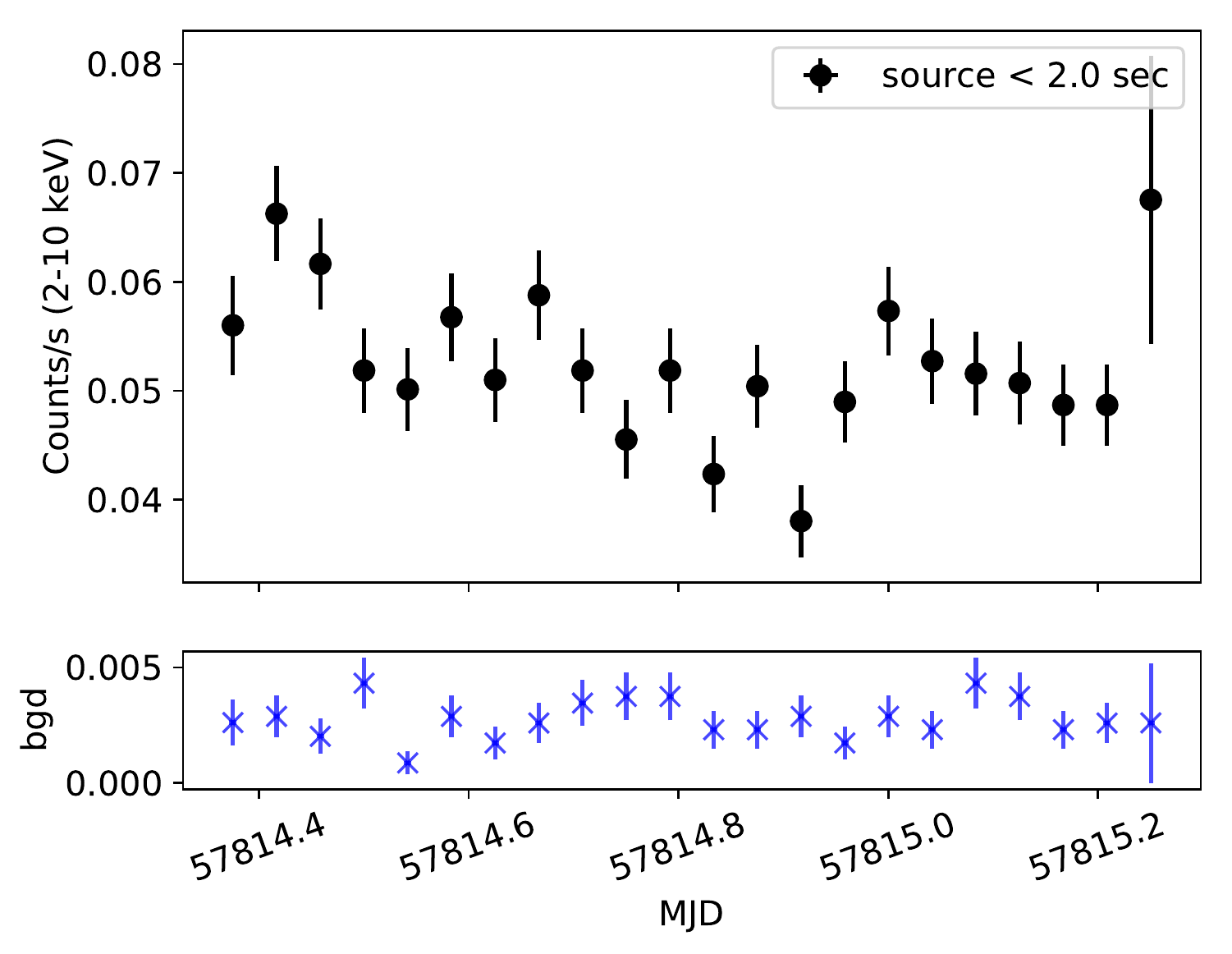}
\label{fig:chan_2017_h}}
\subfigure[Chandra knot A (2017-03-02)]{
\includegraphics[width=7cm,clip]{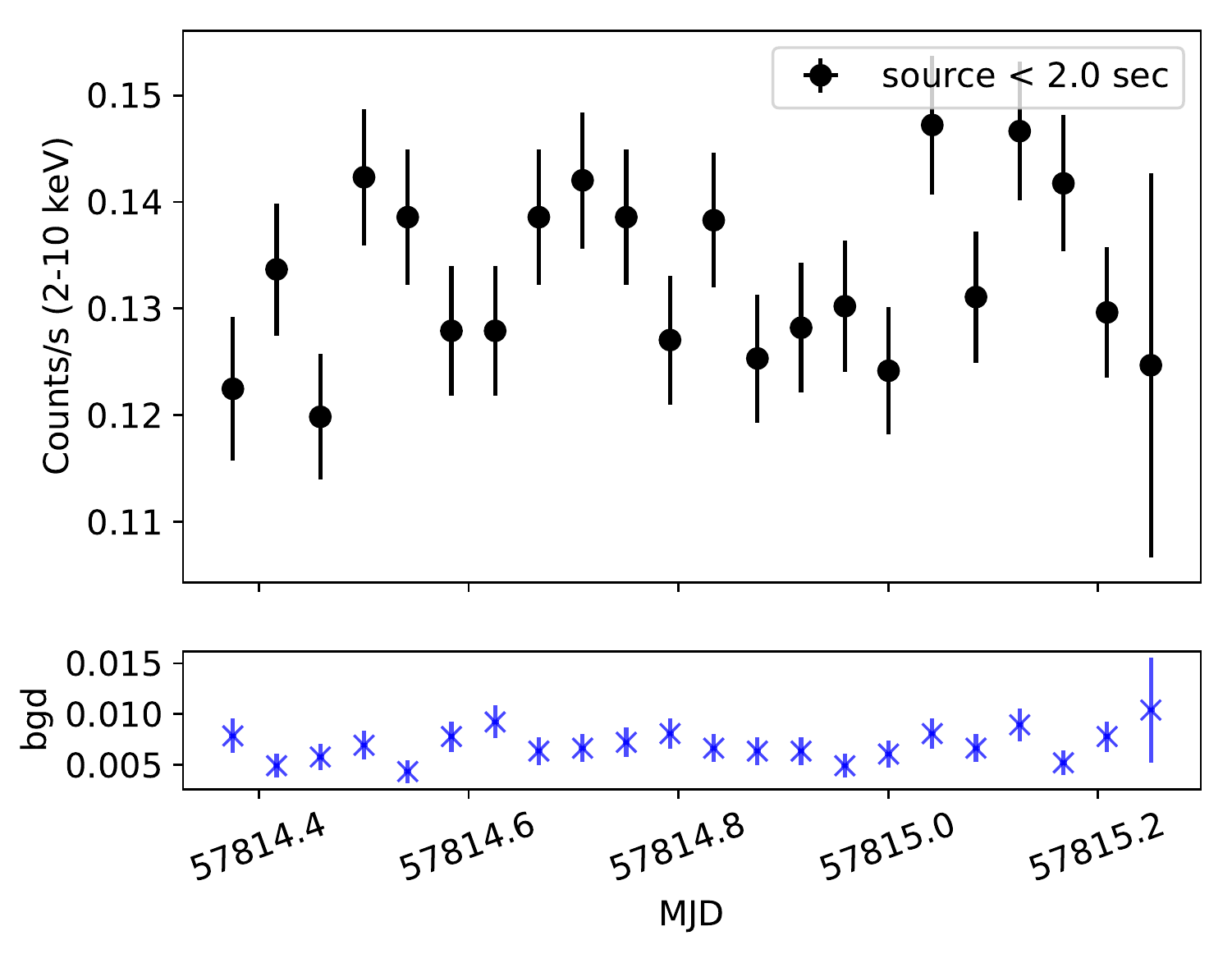}
        \label{fig:chan_2017_k2}}
\\
\subfigure[NuSTAR (2017-02-15)]{
\includegraphics[width=7cm,clip]{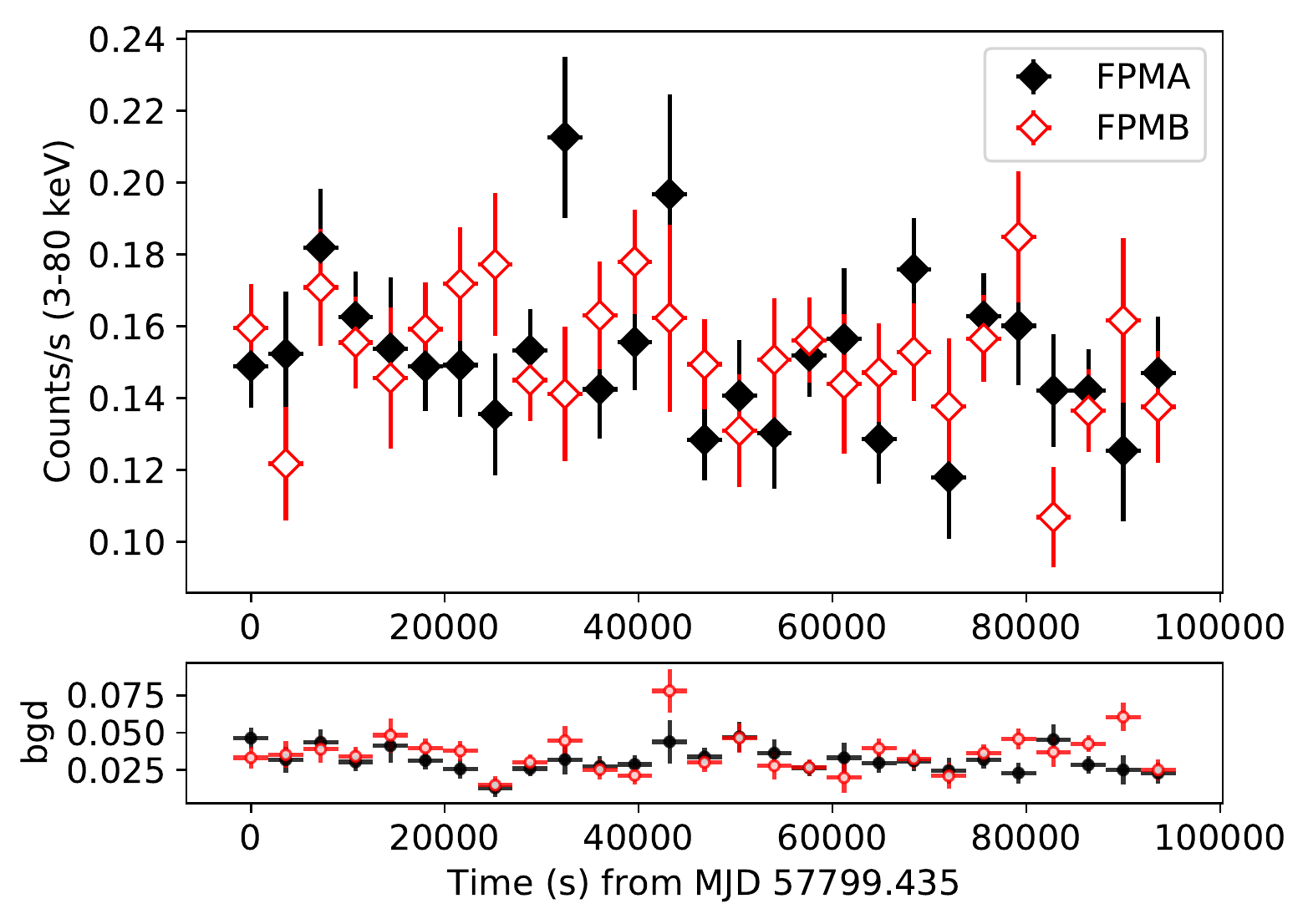}
\label{fig:lc_57799}
        }
                \subfigure[Suzaku (2006-11-29)]{
\includegraphics[width=7cm,clip]{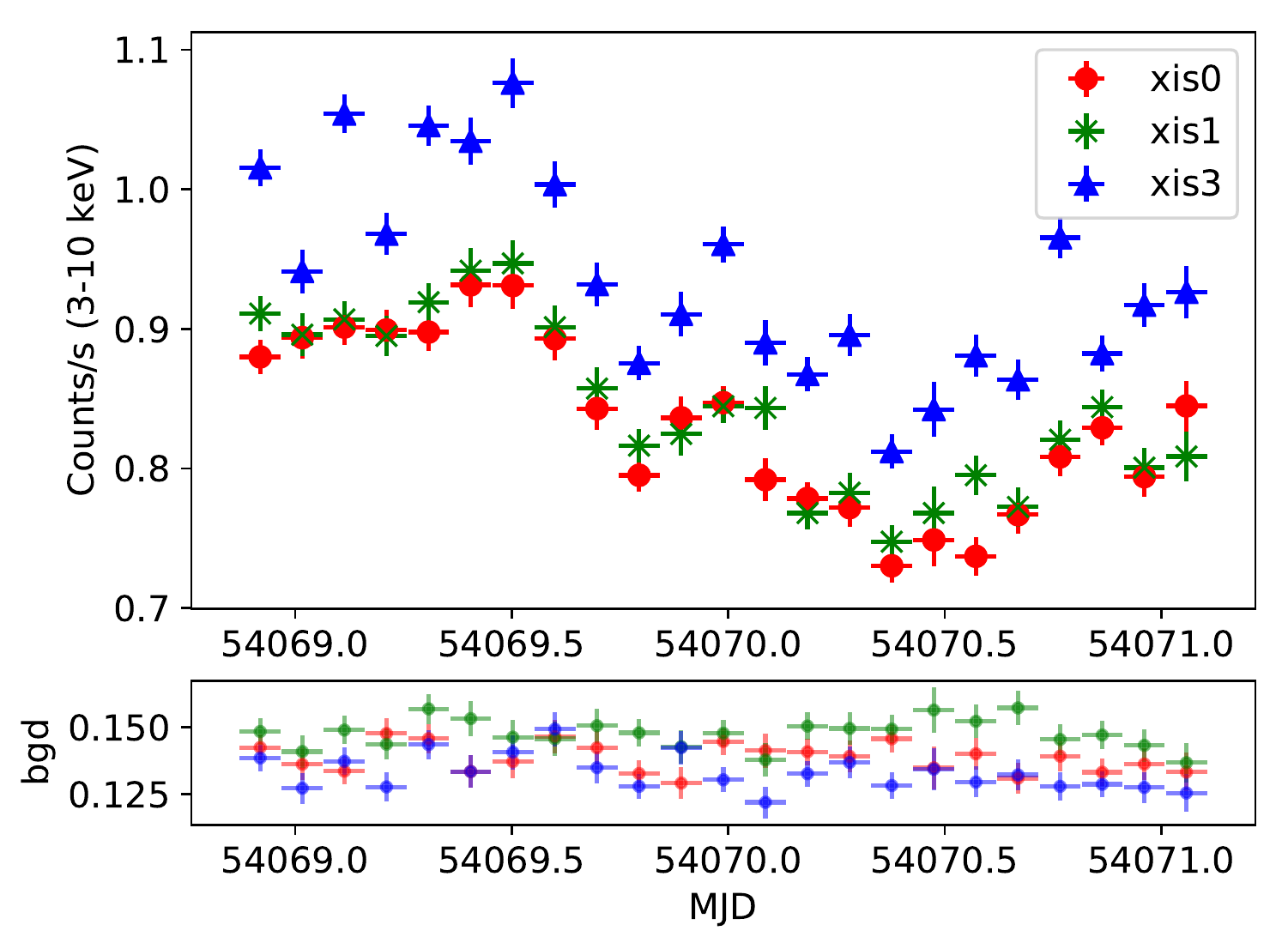}
\label{fig:suza_2006}}\\
\end{tabular}
\caption{(a)-(d) Light curves of each observation of core and HST-1 by Chandra in 2.0--10.0 keV with a 3600 sec per 1 bin. The upper and lower light curve is for the source region and background region, respectively. Background light curves show variability from the mean value.
(e) Light curve of NuSTAR in 3.0--80.0 keV with 3600 sec per 1 bin.
The red and black data points represent FPMA and FPMB, respectively.
(f) Light curve of Suzaku in 3.0--10.0 keV with 8400 sec per 1 bin.
The red, green, and blue data points represent XIS0, XIS1, and XIS3, respectively.}
\label{fig:all_lc}
\end{figure*}

To evaluate a short-term time variability, we calculated a RMS from X-ray light curves of each observation. The RMS is defined as the following equation \citep{Shirai08}:

\begin{equation}
\mathrm{RMS} = \sqrt{\frac{\sum_i [(d_i -\bar{d})^2-e_i^2]}{N\bar{d}^2}}
\label{eq:rms}
\end{equation}

where $d_i$ is a count rate of $i$th data,  $N$ is the number of data points, $\bar{d}$ is an average, and $e_i$ is a statistical error of $d_i$.
Table \ref{tbl:obs} summarizes the RMS values of each observation.
The RMS of observations with $\sum_i [(d_i - \bar{d})^2-e_i^2] < 0$ is not shown, indicating no variability.
As a result, some data showed significant variability.
Figure \ref{fig:all_lc} shows light curves only for observations with a nonzero RMS with 2$\sigma$ significance.

Intraday variability is seen in Suzaku observation in 2006 (Fig.\ref{fig:suza_2006})  and Chandra observation in 2017 (ID:18612, Fig.\ref{fig:chan_2017}).
For the Suzaku light curve, count rate changes by approximately 30\% with a time scale of 0.3 days.
There is a difference in count rates between XIS0 and XIS3, which is attributable to the off-axis observation \citep{Suzaku15}.
The Chandra light curve of the core in 2017 showed approximately 40\% variability with a time scale of 0.5 days.
However, since the extraction region of Chandra data was as small as 2.0 arcsec, the fluctuation of the satellite attitude may cause a variability of the count rate.
Therefore, we made light curves with different extraction radii of 0.4, 1.0, 2.0, and 4.0 arcsec to evaluate the systematic effects.
From this analysis, RMS became 0.10$\pm$0.01, 0.09$\pm$0.01, 0.08$\pm$0.01, and 0.06$\pm$0.01, for 0.4, 1.0, 2.0, and 4.0 arcsec, respectively.
The RMS decreased as the radius increased, but variability was significant, even for a large radius.
Fig.\ref{fig:chan_2017_h} shows a significant variability of HST-1 in 2017. However, the variability was quite similar to that of the core, and thus the variability was due to the core signals.
The light curves of knot A in 2016 and 2017 showed an insignificant variability pattern (Fig.\ref{fig:chan_2017_k1} and \ref{fig:chan_2017_k2}).
The NuSTAR light curve in 2016 showed an insignificant variability pattern because two instruments showed a different variability pattern (Fig. \ref{fig:lc_57799}).\\

Therefore, variability was detected from the Suzaku X-ray observation data  in  2006  and Chandra observation of the core data in 2017.

For Suzaku and Chandra data with significant variability, we analyzed the X-ray spectra as described in Section 2 to obtain a photon index and flux of the power-law component. As a result, a power-law photon index became $2.38^{+0.07}_{-0.04}$ and $1.96^{+0.05}_{-0.04}$, and a flux in 2.0--10.0 keV became $2.71^{+0.18}_{-0.18} \times 10^{-11}$ erg cm$^{-2}$ s$^{-1}$ and $1.86^{+0.17}_{-0.16} \times 10^{-12}$ erg cm$^{-2}$ s$^{-1}$ for Suzaku in 2006 and Chandra core in 2017, respectively.
These fluxes were consistent with a long-term light curve of core and HST-1 \citep{Yang_2019}.
The photon index was different between 2006 and 2017, inferring a different emission mechanism.

\begin{figure}
 \centering
 \includegraphics[width=8cm,clip]{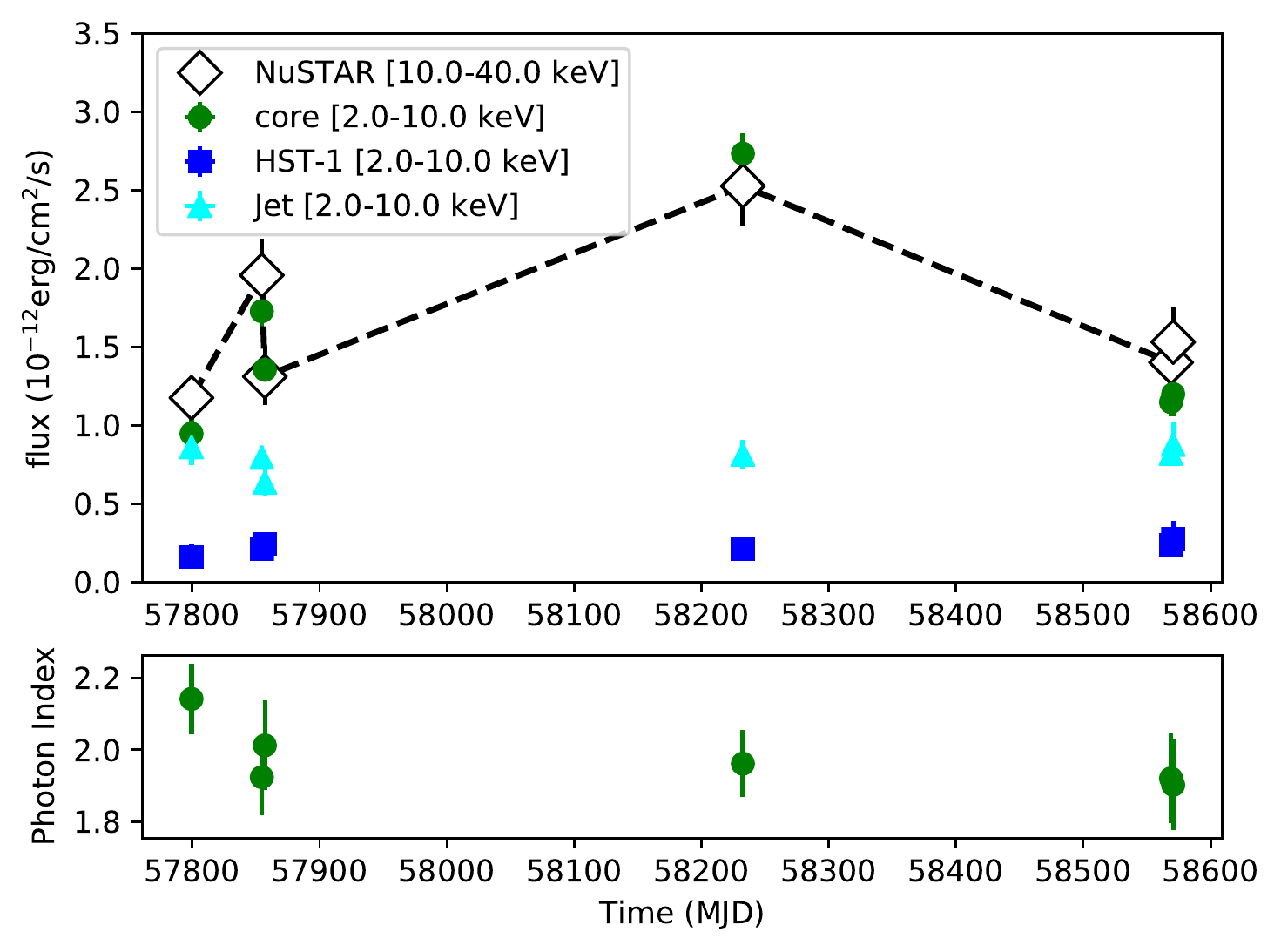}
        \caption{The upper panel shows X-ray light curves of M87 from 2017 to 2019. The lower panel shows a time history of the core photon index of the core obtained from the simultaneous spectral fitting of Chandra and NuSTAR data.
The white diamonds represent flux in 10.0--40.0 keV from only the NuSTAR spectra, and the dark green filled circles, blue-filled squares, and light blue-filled squares are flux in 2.0--10.0 keV of the core, HST-1, and jet, respectively, obtained by the simultaneous fitting of Chandra and NuSTAR spectra.}
 \label{fig:nu_lc}
\end{figure}

We did not find any short-term significant variability for NuSTAR data, but found  that  one  of the six  observations  showed  a  very  different  X-ray  flux.  
As described in $\S$ 2.2.1, we fitted Chandra and NuSTAR spectra simultaneously or only the NuSTAR spectrum of each observation, and obtained a power-law photon index and flux.
Figure \ref{fig:nu_lc} shows the 10.0--40.0 keV X-ray light curves obtained from the NuSTAR spectrum and the 2.0--10.0 keV light curves for the core, HST-1, and jet obtained with a simultaneous fitting of Chandra and NuSTAR spectra.
Remarkably, the flux in 2018 was 2.5 times higher than other observations. 
Chandra core flux was also higher in the same observation, but the fluxes of HST-1 and jet were almost constant.
\cite{Wong_2017} reported a flux of $7.7_{-1.0}^{+1.1} \times 10^{-13}$ erg cm$^{-2}$ s$^{-1}$ and a photon index of $2.12_{-0.13}^{+0.12}$ (20.0--40.0 keV) for the NuSTAR data in 2017.
Our result of the flux of $9.1_{-1.3}^{+1.4} \times 10^{-13}$ erg cm$^{-2}$ s$^{-1}$  and the photon index of = $2.20_{-0.11}^{+0.11}$ (20.0--40.0 keV) were consistent with those of \citet{Wong_2017}\\
As described in \S2, the Chandra core in the brightest state in 2018 was affected by pile-up.
The pile-up fraction is around 7.5\%, and thus the flux could be reduced by at most 7.5\%.
In addition, this could make the photon index smaller by $\sim$0.2, based on the Chandra ABC guide to Pileup
\footnote{ https://cxc.harvard.edu/ciao/download/doc/pileup\_abc.pdf },
However, these effects are within the statistical errors, and thus our result and discussion do not change significantly.

\section{Discussion}
\label{chap_dis}

The results obtained in this study are summarized below.
First, the Suzaku observation in 2006 and the Chandra core observations in 2017 showed significant intraday X-ray variability.
In the Suzaku light curve, the variability amplitude was approximately 30\% with a time scale of about 0.3 days.
For Chandra, the amplitude was approximately 40\% with a time scale of 0.5 days.
In 2006, the HST-1 was much brighter than the core in the X-ray band \citep{Yang_2019}.
Moreover, in 2017 the core was brighter than the HST-1. 
The spectrum obtained by Suzaku in 2006 showed a steep photon index of 2.38$^{+0.07}_{-0.04}$, and the core obtained by Chandra and NuSTAR in 2017 had a hard photon index of $1.96^{+0.05}_{-0.04}$.
The minimum size $R_{\rm{min}}$ of the emission region was estimated by the X-ray variability timescale $\Delta t$ as follows:

\begin{equation}
R_{\rm{min}} = \frac{c\Delta t\delta_D}{1+z}
  \label{eq:lag1}
\end{equation}
where $\delta_D$ is a Doppler factor and $z$ is a redshift. We found $\Delta t \sim$0.3 day in 2006, and thus we obtain $R_{\rm{min}}$ $\simeq$ 1.0$ \delta_D \times$10$^{15}$ cm.
Here, a central black hole mass was assumed to be $M_{\rm{BH}}$ = (6.5$\pm$0.7)$\times$ $10^{9}$ $M_\odot$\citep{EHT2019},
The Schwarzschild radius is $R_s \sim 2.0 \times 10^{15} \rm{cm}$.
Thus, the size of the emission region was as compact as about $\sim R_{\rm{s}}$.
Suzaku/XIS cannot resolve each component of the M87 jet, but knots D and A were quite fainter than the core and HST-1.
In addition, X-ray spectra of knots D and A were reported to be soft with a photon index of 2.4-2.6, $\simeq$ 2.2, respectively \citep{xiao18} and thus it is difficult for the knots D and A to explain a rapid variability above 3.0 keV.
\noindent \\

From the result of Chandra on 2006 November 13, a count rate ratio of HST-1 to the core was approximately about 4.45 \citep{Yang_2019}.
If the variability was only due to the core, a count rate ratio of the core must vary with an amplitude of $\sim 260\%$.
Such an extremely large amplitude of variability has never been observed for the M87 core, and thus it is unlikely.

Hubble Space Telescope observed M87 on 2006 May 23, and November 28.
The ultraviolet flux of the core and HST-1 increased by a factor of 1.7 and 1.4, respectively, from May to November
\citep{Madrid09}.
Notably, Suzaku  observation  was performed in 2006 November.
Radio observations were performed by VLBI in 2006 May, 2007 January, and 2007 April, and flux ratio of HST-1 in January was higher by a factor of 1.5 than that in 2006 May, but it decreased in 2007 April \citep{Chang10}.
Moreover, other components of the M87 did not show variability, indicating
that HST-1 was active in 2006 and likely the origin of intraday X-ray variability.
The X-ray spectrum in 2006 was steep with a photon index of 2.38.
\cite{Jong2015} showed that X-ray emission observed by Suzaku could be explained by a high-energy tail of synchrotron, based on the spectral energy distribution (SED) modeling and thus concluded that HST-1 is likely the origin of TeV gamma-rays.
These results implied that X-ray intraday variability in 2006 observed by Suzaku/XIS can be explained by synchrotron radiation from HST-1.

Assuming that X-ray decay was due to electron cooling, we could estimate a magnetic field strength.
The cooling time scale was expressed by the following equation from \cite{Tashiro95}
\begin{equation}
\tau_s = 3.2\times10^4 \times B^{-3/2} \times E_{\rm{ph}}^{-1/2} \times \delta^{-1/2} (\rm{s})
\end{equation}
where, $\tau_s$ is the decay time scale, $B$ is the strength of magnetic field, $E_{\rm{ph}}$ is the photon energy in the observer frame, and $\delta$ is the Doppler factor.
Using a decay time scale of 0.3 days, and $E_{\rm{ph}} = 6500$ eV, and a Doppler factor $\delta$, the magnetic field became $B$ = 1.94$\delta^{1/3}$ mG.
This value was consistent with that in \cite{Park19}, which was based on the study of Faraday rotation.
However, it was quite smaller than 0.6 G obtained by \cite{Jong2015}, which was based on the SED modeling of HST-1.
This inconsistency might be due to the
SED modeling in \cite{Jong2015} using TeV gamma-ray
ﬂare data between 2003 and 2006. If the magnetic ﬁeld was
increased in only the TeV gamma-ray range, by shock
compression, our result could be explained.
Using the estimated magnetic field, electron energy emitting synchrotron X-ray emission should be $8.7\delta^{-1/6}$ TeV, 
and thus particle acceleration up to TeV was indicated to occur in the HST-1.

\cite{Hardcastle_2015} mentioned that the hotspot of radio galaxy Pictor A, which was 160 kpc from the core showed yearly variability.
Although the hotspot of Pictor A was further away from the core than HST-1, indicating that particle acceleration could occur far away at sub-kpc or kpc from the central black hole.
Such particle acceleration could be in the recollimation zone of the jet \citep{Stawarz06}.

What about an X-ray variability of the core with a timescale of $\Delta t\sim$0.5 days in 2017 is?
VERITAS observations in 2008 showed that, the X-ray flux of the core simultaneously varied with TeV gamma-ray flux, suggesting that the core X-ray emission originated from the jet as well as TeV gamma-ray \citep{Acciari_2008}.
Therefore, the 2017 variability could be due to jet emission at the core.
The power-law photon index of the core X-ray spectrum in 2017 is 1.96$^{+0.05}_{-0.04}$, which could be the inverse Compton scattering component rather than synchrotron radiation if it comes from the jet.
Other possibilities of the origin of harder X-ray emission  besides the jet include an advection dominated accretion flow (ADAF), which  had  been highlighted in previous studies (e.g. \cite{Harris2011} and \cite{Yuan_2014}).
Let us consider whether the observed variability time scale of 0.5 day in 2017 can be explained by ADAF. It is believed that X-ray emission of low luminosity AGN like M81 comes from ADAF.
The mass of the supermassive black hole in M81 was $7.0^{+2}_{-1}\times10^{7} M_{\odot}$ \citep{Schodel07}, and the X-ray variability time scale was confirmed to be about one day \citep{Parola2003}.
Therefore, in the case of M81, the size of the radiation region is 25 $R_{s}$, which is much larger than $\sim$ $R_{s}$ of M87 estimated in this study.
The origin of X-ray emission of M81 was investigated by \cite{Quataert99}
who modeled the multiwavelength SED. They concluded that the ADAF model is promising.
The size of 25 $R_s$ for the emission region was consistent with the ADAF view.
In other words, the intraday variability of the M87 core leads to the emission region size of $\sim R_s$ and thus ADAF origin is unlikely.
Accordingly, it is inferred that the intraday X-ray variability of the core in 2017 is due to the jet emission.
Intraday variability of the core implies a smaller emission region than
the photon ring ($6R_s$) \citep{EHT2019}. Future observations by the Event Horizon Telescope may reveal a radio variability
of the photon ring region.

In the six NuSTAR and Chandra observations in 2017-2019, the flux of the core was high by a factor of 2.5 in only one observation of 2018 April.
The core emission in this period is considered as jet inverse Compton or ADAF radiation rather than jet synchrotron radiation, as discussed above.
Since the variability time scale is just constrained to be longer than 1 day, we cannot discuss the origin of variability any further.\\

\section{Conclusion}
\label{chap_con}

We searched intraday X-ray variability of M87 from Chandra, NuSTAR, and Suzaku archival data.
As a result, Suzaku observation in 2006 and Chandra core observation in 2017 showed significant intraday variability with a time scale of 0.3 days and 0.5 days, respectively.
A much brighter X-ray flux of HST-1 than the core in 2006 indicates that the variability comes from HST-1.

The X-ray spectrum of M87 in 2006 was steep with a photon index of $2.38^{+0.07}_{-0.04}$, but the core in 2017 was hard with a photon index of $1.96^{+0.05}_{-0.04}$.
The steep spectrum of HST-1 seems to be a high-energy tail of synchrotron emission.
Assuming the decay time scale corresponds to a synchrotron-cooling time scale,
the magnetic field strength is calculated to 1.94$\delta^{1/3}$ mG,
yielding the electron energy of $8.7\delta^{-1/6}$ TeV from the below equation\citep{Kifune}.
\begin{equation}
    \gamma E_{e} = \gamma^2 m_ec^2 = \sqrt{(B_{cr}/B)\times(E_{ph}/m_e c^2)}m_ec^2
\end{equation}
where, $E_{e}$ is a electron energy, $\gamma$ and $m_ec^2$ is a Lorentz factor and rest mass of the electron, respectively, and $B_{cr}$ is a critical magnetic field ($=4.4 \times 10^{13}$ (G)).

It indicates that particle acceleration up to TeV occurs at subkiloparsec from the central black hole.

The core hard spectrum in 2017 indicates the possibility of ADAF emission or inverse Compton in the jet, but we obtain minimum emission size as compact as the Schwarzschild radius from the variability time scale, and thus ADAF origin is not preferred.
From these results, we suggest that particle acceleration occurs in both the core and subkiloparsec jets in M87. \\

\begin{center}
ACKNOWLEDGEMENTS\\
\end{center}
This research has made use of data and software provided by the High Energy Astrophysics
Science Archive Research Center (HEASARC), which is a service of the Astrophysics Science
Division at NASA/GSFC.

\bibliography{sample631}{}
\bibliographystyle{aasjournal}



\end{document}